\documentclass[twocolumn,showpacs,preprintnumbers,amsmath,amssymb]{revtex4}
\usepackage{graphicx}% Include figure files
\usepackage{dcolumn}% Align table columns on decimal point
\usepackage{bm}% bold math

\begin{document}

%\preprint{APS/123-QED}

\title{Demonstration of Temporal Distinguishability \\in a Four-Photon State and a Six-Photon State}
% Force line breaks with \\

\author{G. Y. Xiang$^1$, Y. F. Huang$^1$, F. W. Sun$^1$, P. Zhang$^1$, Z. Y. Ou$^{1,2}$, and G. C. Guo$^1$}
 \affiliation{$^1$Key Laboratory of Quantum Information,
 University of Science and Technology of China, \\CAS, Hefei, 230026, the People's Republic of China
 \\$^2$Department of Physics, Indiana
University-Purdue University Indianapolis, 402 N. Blackford
Street, Indianapolis, IN 46202, USA}

\date{\today}% It is always \today, today,
             %  but any date may be explicitly specified

\begin{abstract}
An experiment is performed to demonstrate the temporal
distinguishability of a four-photon state and a six-photon state,
both from parametric down-conversion. The experiment is based on a
multi-photon interference scheme in a recent discovered NOON-state
projection measurement. By measuring the visibility of the
interference dip, we can distinguish the various scenarios in the
temporal distribution of the pairs and thus quantitatively
determine the degree of temporal (in)distinguishability of a
multi-photon state.

\end{abstract}

% PACS, the Physics and Astronomy
                             % Classification Scheme.
%\keywords{Suggested keywords}%Use showkeys class option if keyword
                              %display desired
\maketitle

It has been well-known by now that quantum nonlocality is more
dramatic in multi-particle entanglement \cite{ghz}. It was shown
that the amount of locality violation increases with the number of
particles \cite{mer}. Experimental demonstrations of locality
violation have thus been shifted from the traditional test of
two-photon Bell's inequalities \cite{asp,ou1,shih} to the test of
generalized Bell's inequalities for three or four photons in
various states \cite{pan,how,eib,zha,wal}. While entangled
two-photon states are produced naturally from parametric
down-conversion, generation of three- and four-photon entangled
states has to rely on simultaneous two-pair production in
parametric down-conversion. Since pairs are produced randomly in
parametric down-conversion process, this raises a question, that
is, are the two pairs really in an entangled four-photon state or
they are simply independent uncorrelated two pairs?

This question was first attempted by Ou, Rhee, and Wang
\cite{rhe1,rhe2} in an experiment similar to the famous
Hong-Ou-Mandel experiment \cite{hom} but with two pairs of
photons. Recently, a number of experiments were performed to
further address the problem of photon pair distinguishability in
parametric down-conversion \cite{tsu,ou2,bouw,de}. All the
experimental schemes are more or less some sort of multi-photon
interference (either two-photon or four-photon). More recently, a
new scheme was proposed by Ou \cite{ou3} that relies on a newly
discovered NOON-state projection measurement process \cite
{sun1,res,sun2} to characterize quantitatively the degree of
temporal distinguishability of an N-photon state. When it applies
to photon pairs from parametric down-conversion, it shows various
visibility of multi-photon interference for different scenarios in
the temporal distributions of the photons \cite{ou3}.

In this letter, we wish to report on an experimental
implementation of the NOON-state projection measurement for
characterizing the temporal distinguishability of photon pairs
from parametric down-conversion. We find that the temporal
distinguishability depends on the visibility of multi-photon
interference in NOON state projection. When the pairs are
indistinguishable from each other, we obtain the maximum
visibility which starts to decrease as the pairs begin to separate
from each other and becomes a nonzero minimum when they are well
separated and completely distinguishable.

The key idea in Ref.\cite{ou3} for characterizing temporal
distinguishability is the NOON-state projection measurement, as
depicted in Fig.1 for $N=4$ and 6. This was recently proposed and
demonstrated by Sun {\it et al.} \cite {sun1,sun2} and Resch {\it
et al.} \cite{res} for multi-photon de Broglie wavelength. This
measurement projects an arbitrary two-mode N-photon state of
polarization in the form of
\begin{eqnarray}
|\Psi_N\rangle = \sum_k c_k|N-k\rangle_H
|k\rangle_V,~~~~\label{Psi-N}
\end{eqnarray}
to a NOON state of $|NOON\rangle =
(|N,0\rangle-|0,N\rangle)/\sqrt{2}$. The outcome of this
measurement is proportional to
\begin{eqnarray}
&&P_N \propto |\langle NOON|\Psi_N\rangle|^2.~~~~\label{P-N}
\end{eqnarray}

\begin{figure}[htb]
\begin{center}
\includegraphics[width= 3in]{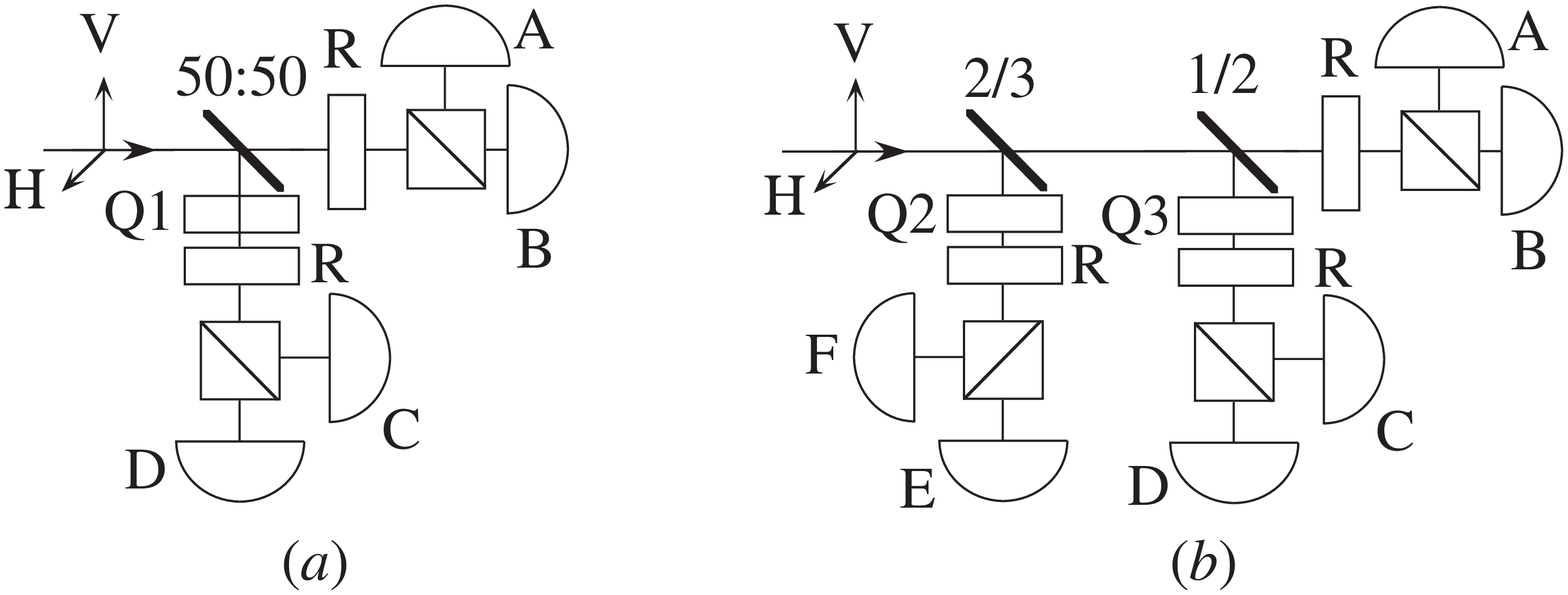}
\end{center}
\caption{\em A NOON-state projection measurement for (a)
four-photon case and (b) six-photon case. Q1 is a phase shifter of
$\pi/2$, Q2 of $2\pi/3$, and Q3 of $4\pi/3$. R is a rotator of
45$^{\circ}$.} \label{fig1}
\end{figure}

The physics behind this projection measurement is an ingenious
arrangement \cite{hof} of beam splitters, phase shifters, and
projection polarizers for the cancellation by destructive
interference of the middle terms in the form of $|N-k\rangle_H
|k\rangle_V$ with $k\ne 0, N$ in Eq.(\ref{Psi-N}). In particular
for $|\Psi_4\rangle = |2\rangle_H|2\rangle_V$ and $|\Psi_6\rangle
= |3\rangle_H|3\rangle_V$, the projection probability in
Eq.(\ref{P-N}) is zero due to orthogonality. These two states are
readily available from Type-II parametric down-conversion with a
quantum state of
\begin{eqnarray}
&&|PDC\rangle = |0\rangle + \eta |1\rangle_H|1\rangle_V
\cr&&\hskip 0.3in+ \sqrt{2}\eta^2 |2\rangle_H|2\rangle_V +
\sqrt{6}\eta^3 |3\rangle_H|3\rangle_V+...,~~~~\label{PDC}
\end{eqnarray}
where $|\eta|^2 << 1$ is the pair production probability.

However, the expression in Eq.(\ref{P-N}) is for single mode
treatment, which means that the four photons or six photons must
be in a single temporal mode (the so called $4\times 1$ or
$6\times 1$ case). So the pairs must be indistinguishable in the
production process. This is normally achieved with an ultra-short
pump pulse for parametric down-conversion so that the time of the
pair production is restricted in the time duration defined by the
pump pulse. But because of the finite duration of the pump pulse,
this is only approximately the case in practice. So the pairs
actually have partial indistinguishability and this will be
reflected in the reduced visibility in the multi-photon
interference in the NOON state projection measurement. As
suggested in Ref.\cite{ou3}, the visibility is then a direct
measure for the degree of indistinguishability.

\begin{figure}[htb]
\begin{center}
\includegraphics[width= 3in]{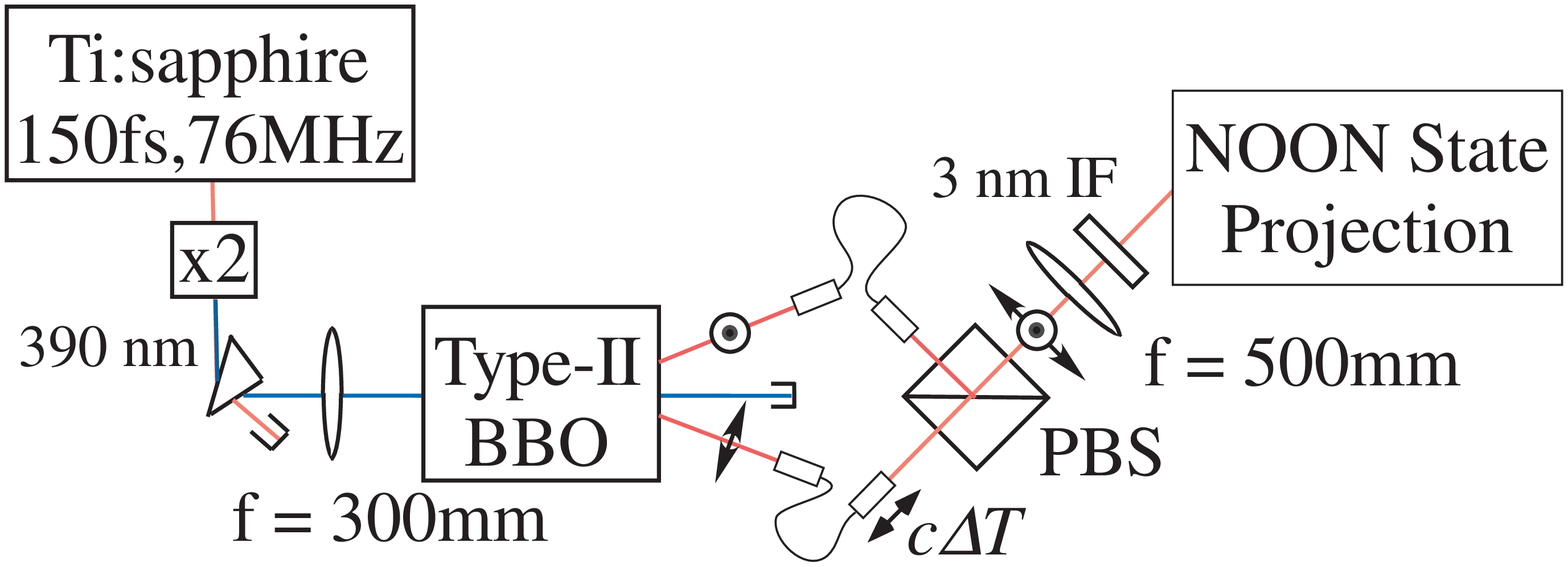}
\end{center}
\caption{\em Sketch of the experimental set-up} \label{fig2}
\end{figure}

The experimental arrangement is shown in Fig.2. A 2-mm long BBO
crystal cut for Type-II parametric down-conversion is pumped at
390 nm by a frequency-doubled femto-second Ti:sapphire laser. The
pump pulse has a width of 150 fsec and a repetition rate of $R_0$
= 76 MHz. The crystal is so oriented that the two conic
down-converted fields (o- and e-rays) at the degenerate frequency
converge into two unidirectional beams \cite{mi}. The two fields
are then coupled to single mode polarization preserving optical
fibers. The outputs of the fibers are directed to a polarization
beam splitter (PBS) to merge into one beam. One of the fiber
outputs is mounted on a translational stage so that we can adjust
the relative delay $c\Delta T$ between the two polarizations. The
recombined beam, after passing through an interference filter of 3
nm width, is sent to the corresponding NOON-state projection
measurement assembly for either four- or six-photon case depicted
in Fig.1. All combinations of two-photon (say, AB, AC, etc.) and
four-photon coincidence (say, ABCD, ACBE, etc.) as well as
six-photon coincidence (ABCDEF) in the six-photon case are
measured as a function of the relative delay $c\Delta T$ between
the two polarizations. Because of the vast amount of coincidence
data and the lack of coincidence units, we measure each
coincidence individually.

\begin{figure}[htb]
\begin{center}
\includegraphics[width= 3in]{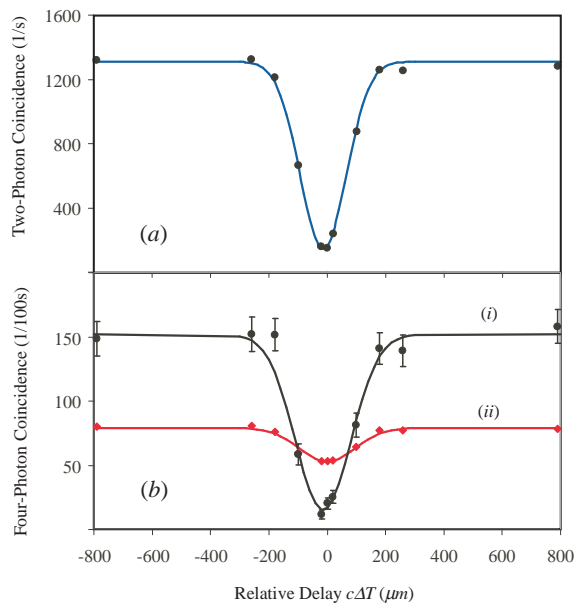}
\end{center}
\caption{\em Measured two-photon coincidences (a) and four-photon
coincidences (b) as a function of relative delay $c\Delta T$. The
circles in (b) are direct measured ABCD coincidences while the
diamonds are indirectly measured coincidences corresponding to the
$2\times 2$ case. The solid curves are Gaussian fit with a
visibility of 90\% and 33\%, respectively.} \label{fig3}
\end{figure}

{\em Four-photon case.} --- When four-photon coincidence is
measured in the four-photon NOON state projection scheme in
Fig.1a, the vacuum and two-photon terms in Eq.(\ref{PDC}) make no
contribution whereas the six-photon term is higher order (it does
produce a background that must be subtracted in data analysis).
The four-photon terms can be thought of as two-pair production.
From the properties of parametric down-conversion, we know that
the two photons within a pair are indistinguishable in time. But
since pair production is random, the two pairs are often generated
at two well separated times and are described by the quantum state
of $|1\rangle_{H1}|1\rangle_{V1}
\otimes|1\rangle_{H2}|1\rangle_{V2}$ (the $2\times 2$ case). Or by
chance, the two pairs may be generated in the same time and become
indistinguishable four photons described by the quantum state of
$|2\rangle_H|2\rangle_V$ (the $4\times 1$ case).

For the four-photon NOON-state projection measurement in Fig.1a,
there are six possible combinations of two-photon coincidence.
They are simply AB, CD, AC, BD, AD, BC. Among them, AB and CD show
the typical two-photon Hong-Ou-Mandel dips \cite{hom} with AB
shown in Fig.3a whereas the rest is flat (not shown). The
visibility of the dip is 89\%. The directly measured four-photon
coincidence ABCD data after background subtraction is shown as
solid circles ($i$) in Fig.3b with a Gaussian fit that has a
visibility of 90\%. The points in diamonds ($ii$) are indirectly
measured four-photon coincidence data corresponding to the case
when the measured four-photon coincidence is from two
well-separated pairs of down-converted photons (the $2\times 2$
case).  The four-photon coincidence in this case is from pairwise
accidental coincidence: there are three possibilities of AB+CD,
AC+BD, and AD+BC. So the four-photon coincidence in $2\times 2$
case can be deduced from the measured two-photon coincidences as
\begin{eqnarray}
&&R_4(ABCD)(2\times 2) = [R_2(AB)R_2(CD) + \cr &&\hskip 0.3 in+
R_2(AC)R_2(BD) + R_2(AD)R_2(BC)]/R_0.~~~~~~~\label{R2x2}
\end{eqnarray}
The solid curve is a Gaussian with a visibility of 33\%.

The less than 100\% visibility (90\%) for the directly measured
four-photon coincidence has two origins. One is from imperfect
spatial mode match due to misalignment. This has already reduced
the two-photon visibility to 89\% in Fig.3a. The other origin is
from non-overlapping between the two detected pairs of photons. In
other word, the two pairs are not completely indistinguishable for
us to treat them as in single temporal mode and to use
Eq.(\ref{P-N}) for four-photon coincidence. Ref.\cite{sun1} has a
complete account of these two effects and derived the visibility
under these imperfect conditions as
\begin{eqnarray}
{\cal V}_4 = {2v_2({\cal A}+3{\cal E})-v_2^2({\cal A+E})\over
3({\cal A+E})}. ~~~~~~\label{VVnn}
\end{eqnarray}
Here $v_2$ is the two-photon visibility from Fig.3a. ${\cal A}$ is
proportional to the absolute square of the four-photon wave
function whereas ${\cal E}(\le {\cal A})$ depends on photon pair
exchange symmetry. When ${\cal E}={\cal A}$, the two pairs are
completely overlapping and the four photons are in an
indistinguishable entangled state described by
$|2\rangle_H|2\rangle_V$ (the $4\times 1$ case). On the other
hand, when ${\cal E} = 0$, the two pairs are completely separated
from each other and become independent (the $2\times 2$ case) with
a four-photon visibility of ${\cal V}_4(2\times 2) = 0.33$ from
Eq.(\ref{VVnn}). This value is exactly the value from the diamond
points in Fig.3b. However, the direct observed four-photon dip in
the circle points in Fig.3b is somewhere in between the two
extreme cases. Substituting the observed values of ${\cal V}_4 =
0.90$ and $v_2=0.89$ in Eq.(\ref{VVnn}) and solving for ${\cal
E}/{\cal A}$, we obtain ${\cal E}/{\cal A} = 0.90$. The quantity
${\cal E}/{\cal A}$ thus provides a measure of partial
indistinguishability between the pairs. The nonzero value of 0.33
for ${\cal V}_4(2\times 2)$ can be thought of as a result of the
indistinguishability between the two photons within each pair. So
the directly measured four-photon dip visibility ${\cal V}_4$ is a
measure of indistinguishability for all the four photons, as
suggested in Ref.\cite{ou3}.

${\cal E}/{\cal A}$ can be independently measured from two-photon
coincidence on one (o- or e-ray) of the the two down-converted
fields \cite{rhe1,rhe2}. This is achieved by blocking one of the
two beams that come to the PBS. The directly measured value is
${\cal E}/{\cal A} = 0.77 \pm 0.06$. Another independent method to
measure ${\cal E}/{\cal A}$ is from the ratio of the values at
infinity delay ($|c\Delta T|=\infty$) in the two data sets in
Fig.3b. Ref.\cite{sun1} gives the ratio as $1+{\cal E}/{\cal A}$
and from Fig.3b, we find ${\cal E}/{\cal A} = 0.92$. This value is
consistent with the one derived from visibility. However, the
value from two-photon coincidence measurement is somewhat smaller
than the ones from four-photon measurement. We believe that this
is caused by spatial mode match between the two pairs of photons.
The four-photon coincidence is more restricted than two-photon
coincidence and thus acts as some sort of spatial mode filtering
resulting in better mode match and higher ${\cal E}/{\cal A}$
values.

\begin{figure}[htb]
\begin{center}
\includegraphics[width= 3in]{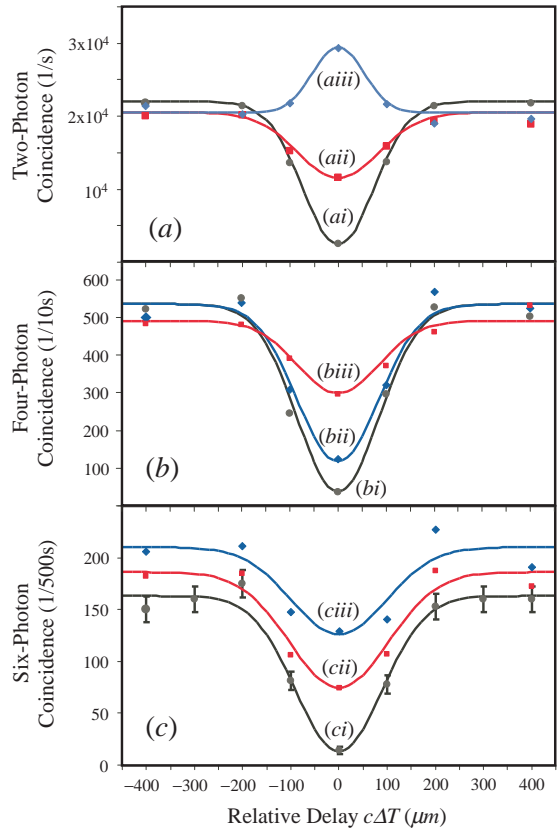}
\end{center}
\caption{\em Measured two-photon coincidences (a), four-photon
coincidences (b), and six-photon coincidences (c) as a function of
relative delay $c\Delta T$. (ai) is for AB, (aii) for AE, and
(aiii) for BE in (a). (bi) is for ABCE, (bii) for ABCD, and (biii)
for BCDE in (b). The solid circles (ci) in (c) are directly
measured ABCDEF coincidence while the diamonds (ciii) are
indirectly measured coincidence corresponding to $2\times 3$ case
and the square points (cii) to the $4\times 1 +2$ case. The solid
curves are Gaussian fit with a visibility of 92\%, 59\%, and 39\%,
respectively. The data in (cii) and (ciii) are multiplied by 2 and
8 respectively to bring them to the same scale as (ci).}
\label{fig4}
\end{figure}

{\em Six-photon case.} --- With six-photon coincidence, the terms
with less than six photons in Eq.(\ref{PDC}) have no contribution.
Since the six photons are from three pairs of down-converted
photons, there are three extreme cases: (1) the three pairs are
generated in the same time and are indistinguishable in the
quantum state of $|3\rangle_H|3\rangle_V$ (the $6\times 1$ case);
(2) two of them are indistinguishable but well separated from the
third pair; they are in the quantum state of
$|2\rangle_{H1}|2\rangle_{V1}\otimes |1\rangle_{H2}|1\rangle_{V2}$
(the $4\times 1+2$ case); (3) all three pairs are well separated
from each other and are in
$|1\rangle_{H1}|1\rangle_{V1}\otimes|1\rangle_{H2}|1\rangle_{V2}
\otimes|1\rangle_{H3}|1\rangle_{V3}$ (the $2\times 3$ case). The
three cases give three different results in the six-photon
NOON-state measurement \cite{ou3}.

In the NOON-state projection for six-photon (Fig.1b), there are 15
different combinations of two-photon or four-photon coincidence.
Among the two-photon coincidences, AB, CD, and EF are the same and
show a typical Hong-Ou-Mandel dip with 100\% visibility in the
ideal case; AC, AE, BD, BF, CE, DF are the same with a dip of an
ideal 50\% visibility; AD, BC, CF, DE, AF, BE are the same with a
bump of an ideal 50\% visibility. Fig.4a shows AB, AC, and AD.
Among the four-photon coincidence, ABCE, ABDF, CDBF, CDAE, EFBD,
EFAC are the same with a dip of 100\% visibility ideally; ABCF,
ABDE, CDBE, CDAF, EFBC, EFAD are the same with a dip of 1/3
visibility ideally; ABCD, ABEF, CDEF are the same with a dip of
5/6 visibility ideally. They are plotted in Fig.4b after
background subtraction. The fitted curves give dips with
visibility smaller than the ideal ones. The directly measured
six-photon coincidence data is presented in Fig.4c as solid
circles. The dip in the fitted Gaussian curve has a visibility of
0.92, as compared to the ideal 100\%. The diamond points and
square points are indirectly measured six-photon coincidence
corresponding to the $2\times 3$ (x8) and the $4\times 1+2$ case
(x2), respectively. They are deduced from
\begin{eqnarray}
&&R_6 = \sum_P R_2(P)R_4(P)/R_0,~~~~~~~~~\label{R6}
\end{eqnarray}
where $P$s are the 15 different combinations of (AB) (CDEF). For
the $4\times 1+2$ case (square points), the quantities
$R_4(CDEF)$, etc. are directly measured four-photon coincidences
(shown in Fig.4b) but for the $2\times 3$ case (diamond points),
they are derived from two-photon coincidences by using a formula
similar to Eq.(\ref{R2x2}). The diamond points and the square
points are fitted to Gaussian functions with visibility of ${\cal
V}_6(2\times 3) = 0.39$ and ${\cal V}_6(4\times 1+2) = 0.59$,
respectively. The ideal values are ${\cal V}_6^{(0)}(2\times 3) =
2/5$ and ${\cal V}_6^{(0)}(4\times 1+2) =3/5$ \cite{ou3}.

The observed visibilities in both $6\times 1$ and $4\times 1+2$
case are close to the ideal values, indicating that the pairs are
almost indistinguishable. This is reflected in the directly
measured ${\cal E/A} =0.91\pm 0.04$ from two-photon coincidence
and the ${\cal E/A}$ value would be even closer to 1 for
six-photon coincidence. Of course, the observed visibility in
$2\times 3$ case is always the predicted value because the two
photons in each pair are truly indistinguishable and we have a
genuine $2\times 3$ case.

In summary, we used the newly discovered NOON-state projection
measurement technique to quantitatively characterize the temporal
distinguishability of a four- or six-photon state. We find the
visibility of the multi-photon interference can be used to
distinguish different scenarios in the temporal distribution of
the photons.

\begin{acknowledgments}
This work was funded by National Fundamental Research Program of
China (2001CB309300), the Innovation funds from Chinese Academy of
Sciences, and National Natural Science Foundation of China (Grant
No. 60121503 and No. 10404027)). ZYO is also supported by the US
National Science Foundation under Grant No. 0245421 and No.
0427647.
\end{acknowledgments}

\end{document}